\documentclass{sig-alternate}
\usepackage{amsmath,amssymb,amscd,epsfig,amsfonts,rotating}
\usepackage{graphicx}
\usepackage{epsfig}
\usepackage{epstopdf}
\usepackage{cite}

  
\begin{document}

\title{Learning the Gain Values and Discount Factors of DCG}
\numberofauthors{3}
\author{
\alignauthor Ke Zhou\\
       \affaddr{Dept. of Computer Science and Engineering}\\
       \affaddr{Shanghai Jiao-Tong University}\\
       \affaddr{No. 800, Dongchuan Road, Shanghai, China 200240}
       \email{zhouke\\@apex.sjtu.edu.cn}
\alignauthor Hongyuan Zha\\
       \affaddr{College of Computing}\\
       \affaddr{Georgia Institute of Technology}\\
       \affaddr{Atlanta, GA 30032}\\
       \email{zha@cc.gatech.edu}
\alignauthor Gui-Rong Xue, Yong Yu\\
       \affaddr{Dept. of Computer Science and Engineering}\\
       \affaddr{Shanghai Jiao-Tong University}\\
       \affaddr{No. 800, Dongchuan Road, Shanghai, China 200240}\\
       \email{\{grxue, yyu\}@apex.sjtu.edu.cn}
}
\maketitle

\begin{abstract}
Evaluation metrics are an essential part of a ranking system, and in the past many evaluation metrics have been proposed in information retrieval and Web search. Discounted Cumulated Gains (DCG) has emerged as one of the evaluation metrics widely adopted for evaluating the performance of ranking functions used in Web search. However, the two sets of parameters, gain values and discount factors, used in DCG are determined in a rather ad-hoc way.
In this paper we first show that DCG is generally not coherent, meaning that comparing the performance of ranking functions using DCG very much depends on the particular gain values and discount factors used. We then propose a novel
methodology that can learn the gain values and discount factors from user preferences over rankings. Numerical simulations illustrate the effectiveness of our proposed methods. Please contact the authors for the full version of this work.
\end{abstract}

\section{Introduction}
Discounted Cumulated Gains (DCG) is a popular evaluation metric for comparing the performance of ranking functions \cite{sigir2000:jarvelin}. It can deal with multi-grade judgments and it also explicitly incorporates the position information of the documents in the result sets through the use of discount factors. However, in the past, the selection of the two sets of parameters, gain values and discount factors, used in DCG is rather arbitrary, and several different sets of values have been used. This is rather an unsatisfactory situation considering the popularity of DCG. In this paper, we address the following two important issues of DCG:

\begin{enumerate}
\item Does the parameter set matter? I.e., do different parameter sets give rise to different preference over the ranking functions?

\item If the answer to the above question is {\it yes}, is there a principled way to selection the set of parameters?
\end{enumerate}

The answer to the first question is {\it yes} if there are more than two grades used in the evaluation. This is generally the case for Web search where multiple grades are used to indicate the degree of relevance of documents with respect to a query. We then propose a principled approach for learning the set of parameters using preferences over different rankings of the documents. As will be shown the resulting optimization problem for the learning the parameters can be solved using quadratic programming very much like what is done in support vector machines for classification. We did several numerical simulations that illustrate the feasibility and effectiveness of the proposed methodology. We want to emphasize that the experimental results are preliminary and limited in its scope because of the use of the simulation data; and experiments using real-world search engine data are being considered.

\section{Related Work}
Cumulated gain based measures such as DCG \cite{sigir2000:jarvelin} have been applied to evaluate information retrieval systems. Despite their popularity, little research has been focused on analyzing the coherence of these measures to the best of our knowledge. The study of \cite{sigir01:voorhees} shows that different gain values of DCG can raise different judgements of ranking lists. In this study, we first prove that the DCG is incoherency and then propose a principled method to learn the DCG parameters as a linear utility function.

Learning to rank attracts a lot of research interests in recent years. Several methods have been developed to learn the ranking function through directly optimization performance metrics such as MAP and DCG \cite{sigir2007:xu, sigir07:yue, icml2005:joachims}. These studies focus on learning a good ranking function with respect to given performance metrics, while the goal of this paper is to analysis coherence of DCG and propose a learning method to determine the parameters of DCG.

As we have mentioned in Section \ref{sct:learning}, DCG can be viewed as a linear utility function. Therefore, the problem of learning DCG is closely related to the problem of learning the utility function. Learning utility function is studied under the name of conjoint analysis by the market science community \cite{book:louviere, jmr1997:carrol}. The goal of conjoint analysis is to model the users' preference over products and infer the features that satisfy the demands of users. Several methods have been proposed to model solve the problem \cite{nips2004:chapelle,ms2005:evgeniou}.

\section{Discounted Cumulated Gains}
We first introduce some notation used in this paper. We are interested in ranking $N$ documents ${\cal X}=\{x_1,\dots, x_N\}$. We assume that we have a finite ordinal label (grade) set ${\cal L}=\{\ell_1,\dots,\ell_L\}$. We assume that $\ell_i$ is preferred over $\ell_{i+1}, i=1,\dots, L-1$. In Web search, for example, we can have $${\cal L}=\{\rm Perfect, Excellent, Good, Fair, Bad\},$$ i.e., $L=5$. A ranking of ${\cal X}$ is a permutation
\[\pi=(\pi(1), \dots, \pi(N)),\]
of $(1,\dots, N)$, i.e., the rank of $x_{\pi(i)}$ under the ranking $\pi$ is $i$.

For each label is associated a {\it gain value} $g_i\equiv g(\ell_i)$, and $g_i, i=1,\dots, L$ constitute
the set of gain values associated with ${\cal L}$. The DCG for $\pi$ with the associated labels is computed as
\[ DCG_{g,K}(\pi) = \sum_{i=1}^K c_i g_{\pi(i)}, \quad K=1,\dots, N,\]
where $c_1 > c_2 > \dots > c_K > 0$ are the so-called {\it discount factors} \cite{sigir2000:jarvelin}.

The gain vector $g=[g_1,\dots, g_L]$ is said to be {\it compatible} if
$g_1 > g_2 > \dots > g_L$. If two gain vectors $g^A$ and $g^B$ are both compatible, then we say they are compatible with each other. In this case, there is a transformation $\phi$ such that
\[ \phi(g^A_i) = g^B_i, \quad i=1,\dots, L,\]
and the transformation $\phi$ is {\it order preserving}, i.e.,
 \[ \phi(g_i) > \phi(g_j), \quad \mbox{\rm if $g_i > g_j$.}\]

\section{Incoherency of DCG}
Now assume there are two rankers $A$ and $B$ using DCG with gain vectors $g^A$ and $g^B$, respectively. We want to investigate how coherent $A$ and $B$ are in evaluating different rankings

\subsection{Good News}
First the good news: if $A$ and $B$ are compatible, then $A$ and $B$ agree on which set of rankings is optimal, i.e., which set of rankings have the highest DCG. We first state the following well-known result.

{\bf Proposition 1.}
Let $a_1\geq \dots \geq a_N$ and $b_1\geq \dots \geq b_N$. Then
\[ \sum_{i=1}^N a_i b_i = \max_{\pi} \; \sum_{i=1}^N a_ib_{\pi(i)}.\]

It follows from the above result that any ranking $\pi$ such that
\[ g_{\pi(1)} \geq g_{\pi(2)} \geq \cdots \geq g_{\pi(K)}\]
achieves the highest $DCG_{g,K}$, as long as the gain vector $g$ is compatible.

How about those rankings that have smaller DCGs? We say two compatible rankers $A$ and $B$ are {\it coherent}, if they score any two rankings coherently, i.e., for rankings $\pi_1$ and $\pi_2$,
$$
DCG_{g^A,K}(\pi_1) \geq DCG_{g^A,K}(\pi_2)
$$
if and only if
$$
DCG_{g^B,K}(\pi_1) \geq DCG_{g^B,K}(\pi_2),
$$
i.e., ranker $A$ thinks $\pi_1$ is better than $\pi_2$ if and only if
ranker $B$ thinks $\pi_1$ is better than $\pi_2$. Now the question is whether compatibility implies coherency. We have the following result.

{\bf Theorem.} If $L=2$, then compatibility implies coherency.

{\bf Proof.} Fix $K>1$, and let \[c=\sum_{i=1}^K c_i.\] When there are only two labels, let the corresponding gains be $g_1,  g_2$. For a ranking $\pi$, define
\[ c_1(\pi) = \sum_{\pi(i)=\ell_1} c_i, \quad c_2(\pi) = c-c_1(\pi). \]
 Then
\[ DCG_{g,K}(\pi) = c_1(\pi) g_1 + c_2(\pi) g_2.\]
For any two rankings $\pi_1$ and $\pi_2$,
$$
DCG_{g^A,K}(\pi_1) \geq DCG_{g^A,K}(\pi_2)
$$
implies that
\[ c_1(\pi_1) g_1^A + c_2(\pi_1) g_2^A > c_1(\pi_1) g_1^A + c_2(\pi_1) g_2^A\]
which gives
\[ (c_1(\pi_1)-c_1(\pi_2))(g_1^A-g_2^A) > 0.\]
Since $A$ and $B$ are compatible, the above implies that
\[ (c_1(\pi_1)-c_1(\pi_2))(g_1^B-g_2^B) > 0.\]
Therefore $DCG_{g^B,K}(\pi_1) \geq DCG_{g^B,K}(\pi_2)$. The proof is completed by exchange $A$ and $B$ in the above arguments.

\subsection{Bad News}
Not too surprisingly, compatibility does not imply coherency when $L>2$. We now present an example.

{\bf Example.} Let ${\cal X} = \{x_1,x_2,x_3\}$, i.e., $N=3$. We consider $DCG_{g, K}$ with $K=2$. Assume the labels of $x_1,x_2,x_3$ are $\ell_2, \ell_1, \ell_3$, and for ranker $A$, the corresponding gains are $2,3,1/2$. The optimal ranking is $(2, 1, 3)$. Consider the following two rankings,
\[ \pi_1 = (1,3,2), \quad \pi_2 = (3,2,1)\]
None of them is optimal. Let the discount factors be
\[ c_1 = 1+\epsilon, \quad c_2 = 1-\epsilon, \quad 1/4 < \epsilon <1.\]
It is easy to check that
\[ DCG_{g^A, 2}(\pi_1) = 2c_1 + (1/2)c_2 >\]
\[ > (1/2)c_1 + 3c_2=DCG_{g^A, 2}(\pi_2).\]
Now let $g^B = \phi(g^A)$, where $\phi(t)=t^k$, and $\phi$ is certainly order preserving, i.e., $A$ and $B$ are compatible. However, it is easy to see that for $k$ large enough, we have
\[ 2^kc_1 + (1/2)^kc_2 < (1/2)^kc_1 + 3^kc_2\]
which is the same as
\[ DCG_{g^B, 2}(\pi_1) < DCG_{g^B, 2}(\pi_2).\]
Therefore, $A$ thinks $\pi_1$ is better than $\pi_2$ while $B$ thinks $\pi_2$ is better than $\pi_1$ even though $A$ and $B$ are compatible. This implies $A$ and $B$ are {\it not} coherent.

\subsection{Remarks}
When we have more than two labels, which is the case for Web search, using $DCG_K$ with $K>1$ to compare the DCGs of various ranking functions will very much depend on the gain vectors used. Different gain vectors can lead to completely different conclusions about the performance of the ranking functions.

The current choice of gain vectors for Web search is rather {\it ad hoc}, and there is no criterion to judge which set of gain vectors are reasonable or natural.


\section{Learning Gain Values and Discount Factors}\label{sct:learning}

DCG can be considered as a simple form of linear {\it utility function}. In this section, we discuss a method to learn the
gain values and discount factors that constitute this utility function.

\subsection{A Binary Representation}
We consider a fixed $K$, and we use
a binary vector $s(\pi)$ of dimension $K\times L$ to represent a ranking $\pi$ considered for $DCG_{g,K}$. Here $L$ is the number of levels of the labels. In particularly, the first $L$-components of $s$ correspond to the first position of the $K$-position ranking in question, and the second $L$-components the second position, and so on.
Within each $L$-components, the $i$-th component is 1 if and only if the item in position one has label $\ell_i, i=1,\dots, L$.

{\sc Example.} In the Web search case, $L=5$, suppose we consider $DCG_{g,3}$, and for a particular ranking $\pi$ the labels of the first three documents are
\[ \mbox{\rm Perfect, Bad, Good}.\]
Then the corresponding $15$-dimensional binary vector $s(\pi)$ is
\[ [1, 0, 0, 0, 0,\;\; 0, 0, 0, 0, 1, \;\; 0, 0, 1, 0, 0].\]
We postulate a utility function $u(s)=w^Ts$ which a linear function of $s$, and $w$ is the weight vector, and we write
\[ w=[w_{1,1},\dots, w_{1,L}, \; w_{2,1},\dots, w_{2,L},\; \dots, w_{k,1},\dots, w_{K,L}].\]
 We distinguish two cases.

{\sc Case 1.} The gain values are position independent. This corresponds to the case
\[ w_{i,j} = c_i g_{j}, \quad, i=1,\dots, K, j=1,\dots, L.\]
This is to say that $c_i, i=1,\dots, K$ are the discount factors, and $g_{j}, j=1,\dots, L$ are the gain values.
It is easy to see that
\[ w^Ts(\pi) = DCG_{g,K}(\pi).\]

{\sc Case 2.} In this framework, we can consider the more general case that the gain values are {\it position dependent}. Then $w_{1,1},\dots, w_{1,L}$ are just the products of the discount factor $c_1$ and the
position dependent gain values for position one, and so on. In this case, there is no need to separate the
gain values and the discount factors. The weights in the weight vector $w$ are what we need.

\subsection{Learning $w$}

We assume we have available a partial set of preferences over the set of all rankings. For example, we can present a pair of rankings $\pi_1$ and $\pi_2$ to a user, and the user prefers $\pi_1$ over $\pi_2$, denoted by $\pi_1 \succ \pi_2$, which translates into $w^Ts(\pi_1) \geq w^Ts(\pi_2)$. Let the set of preferences be
\[ \pi_i \succ \pi_j, \quad (i,j) \in S.\]

In the second case described above, we can formulate the problem as learning the weight vector $w$ subject to a set of constraints
(similar to rank SVM):
\begin{equation}
\min_{w, \; \xi_{ij}} w^Tw + C \sum_{(i,j) \in S} \xi_{ij}^2\label{eqn:obj1}
\end{equation}
subject to
\[ w^T\left(s(\pi_i)-s(\pi_j) \right)\geq 1-\xi_{ij}, \;\; \xi_{ij} \geq 0, \;\;(i,j) \in S.\]
\[ w_{kl}\geq w_{k,l+1}, \;\; k=1,\dots,K, \;\; l=1,\dots, L-1\]

For the first case, we can compute $w$ as in Case 2, and then find $c_i$ and $g_j$ to fit $w$. It is also possible to carry out hypothesis testing to see if the gain values are position dependent or not.

\section{Simulation}
In this section, we report the results of numerical simulations to show the feasibility and effectiveness of the method proposed in Equation (\ref{eqn:obj1}).

\subsection{Experimental Settings}
We use a ground-truth $w$ to obtain preference of ranking lists. Our goal is to investigate whether we can reconstruct $w$ via learning from the preference of ranking lists. The ground-truth $w$ is generated according to the following equation:
\begin{eqnarray}
w_{kl} = \frac{G_l}{\log(k+1)}
\end{eqnarray}
For a comprehensive comparison, we distinguish two settings of $G_l$. Specifically, we set $G_l=l$ in the first setting (Data 1), and $G_l = 2^l-1$ in the second setting (Data 2).

The ranking lists are obtained by randomly permuting a ground-truth ranking list. For example, the ranking lists can be generated by permuting the list $[5, 5, 4, 4, \\3, 3, 2, 2, 1, 1]$ randomly.
We randomly generate different numbers of pairs of ranking lists and use the ground-truth  to judge which ranking lists is preferred. Specifically, if $w^Ts(\pi_1)>w^Ts(\pi_2)$, we have a preference pair $\pi_1\succ\pi_2$. Otherwise we have a preference pair $\pi_2\succ\pi_1$.

\subsection{Evaluation Measures}
Given the estimated $\hat w$ and the ground-truth $w$, we apply two measures to evaluate the quality of the estimated $\hat w$.

The first measure is the precision on a test set. A number of pairs of ranking lists are generated as the test set. We apply $\hat w$  to predict the preference over the test set. Then the precision $\hat w$ is calculated as the proportion of correctly predicted preference in the test set.

The second measure is the similarity of $w$ and $\hat w$. Given the true value of $w$ and the estimation $\hat w$ defined by the above optimization problem, the similarity between $w$ and $\hat w$ can be defined as follows:

\begin{eqnarray}
T(w) &=& (w_{11}-w_{1L}, , \dots, w_{1L}-w_{1L}, \nonumber\\
&&\dots, w_{K1}-w_{KL}, \dots, w_{KL}-w_{KL}) \label{eqn:t}
\end{eqnarray}
We can observe that the transformation $T$ preserve the orders between ranking lists, i.e., $T(w)^Ts(\pi_1) > T(w)^Ts(\pi_2)$ iff $w^Ts(\pi_1) > w^Ts(\pi_2)$. The similarity between $w$ and $\hat w$ is measured by ${\rm sim}(w, \hat w) = \frac{T(w)^TT(\hat w)}{\|T(w)\|\|T(\hat w)\|}$.

\subsection{General Performance}
\begin{figure*}[t]
\begin{minipage}[t]{0.48\textwidth}
\centering
\includegraphics[width=\textwidth]{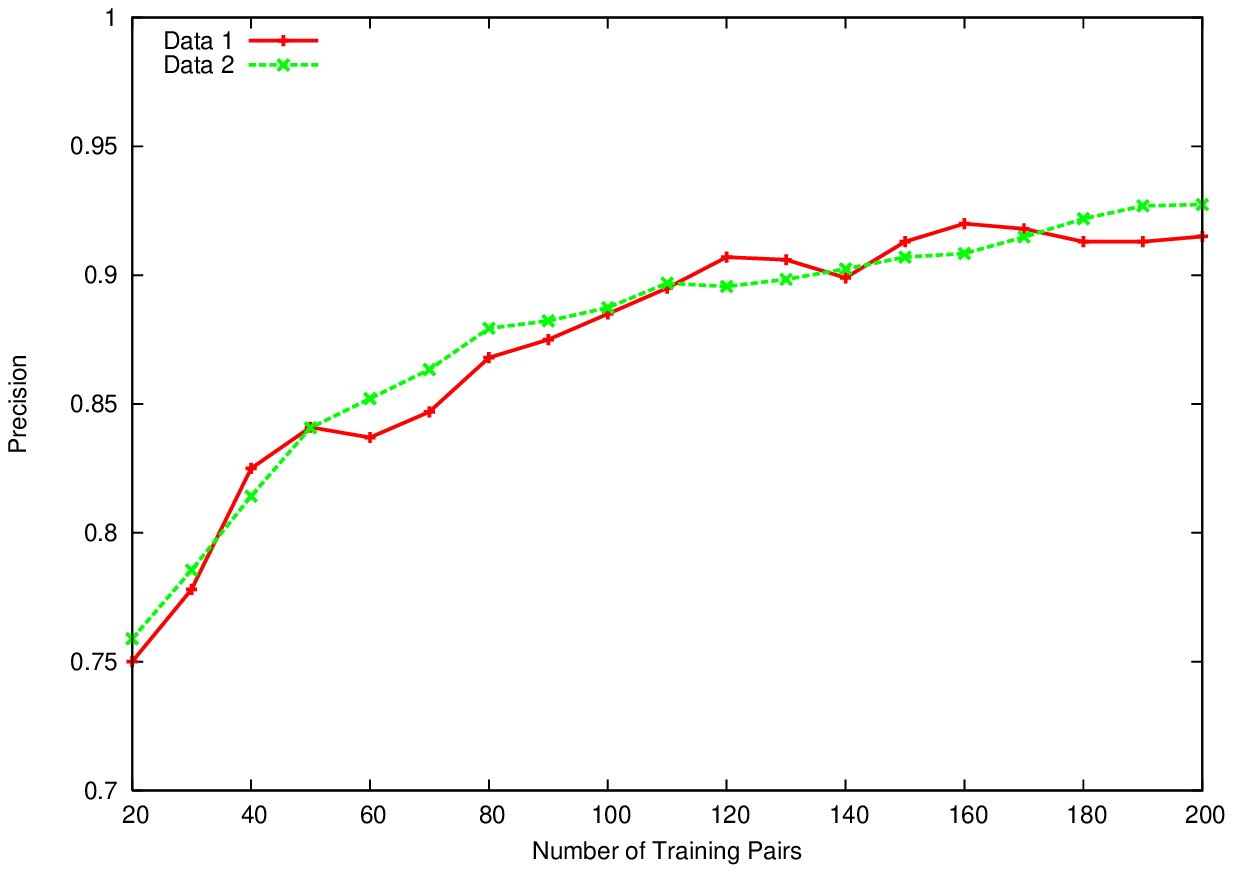}
\caption{Precision over the test set with respect to the number of training pairs} \label{fig:simu_p}
\end{minipage}
\quad
\begin{minipage}[t]{0.48\textwidth}
\centering
\includegraphics[width=1\textwidth]{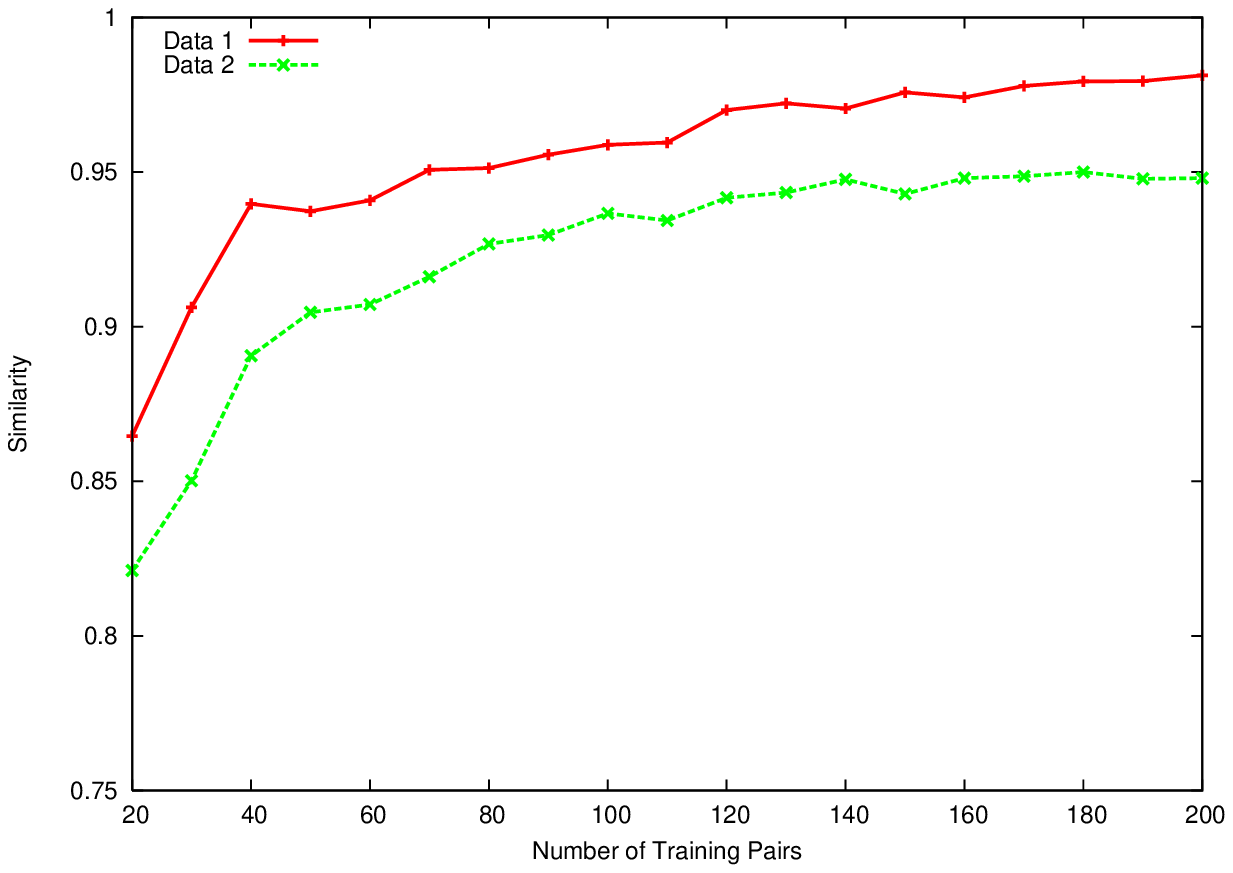}
\caption{Similarity between $w$ and $\hat w$ with respect to the number of training pairs} \label{fig:simu_sim}
\end{minipage}
\end{figure*}

We randomly sample a number of ranking lists and generate preference pairs according to a ground-truth $w$. The number of preference pairs in training set ranges from 20 to 200. We plot the precision and similarity of the estimated $\hat w$ with respect to the number of training pairs in Figure \ref{fig:simu_p} and Figure \ref{fig:simu_sim}. It can be observed from Figure \ref{fig:simu_p} and \ref{fig:simu_sim} that the performance generally grows with the increasing of training pairs, indicting that the preference over ranking lists can be utilized to enhance the estimation of the unity function $w$. Another observation is that the when about 200 preference pairs are included in training set, the precisions in test sets become close to $95\%$ under both settings. This observation suggests that we can estimate $w$ precisely from the preference of ranking lists. We also notice that the similarity and precision sometimes give different conclusions of the relative performance over Data 1 and Data 2. We think it is because the similarity measure is sensitive to the choice of the offset constant. For example,  large offset constants will give similarity very close to 1. Currently, we use $w_{1L}, \dots, w_{KL}$ of the offset constant as in Equation (\ref{eqn:t}). Generally, we refer the precision as a more meaningful evaluation metric and report similarity as a complement to precision.

\begin{figure*}[thb]
\centering
\begin{minipage}[t]{0.48\textwidth}
\includegraphics[width=1.0\textwidth]{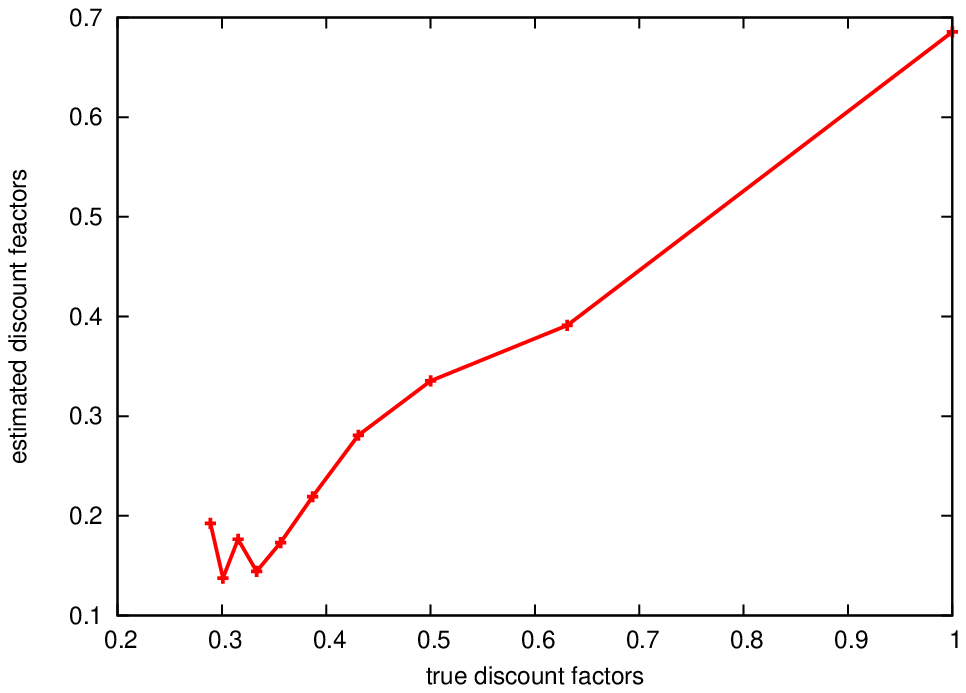}
\caption{Estimated discount factors with respect to true discount factors} \label{fig:rankone_c}
\end{minipage}
\quad
\begin{minipage}[t]{0.48\textwidth}
\center
\includegraphics[width=1.0\textwidth]{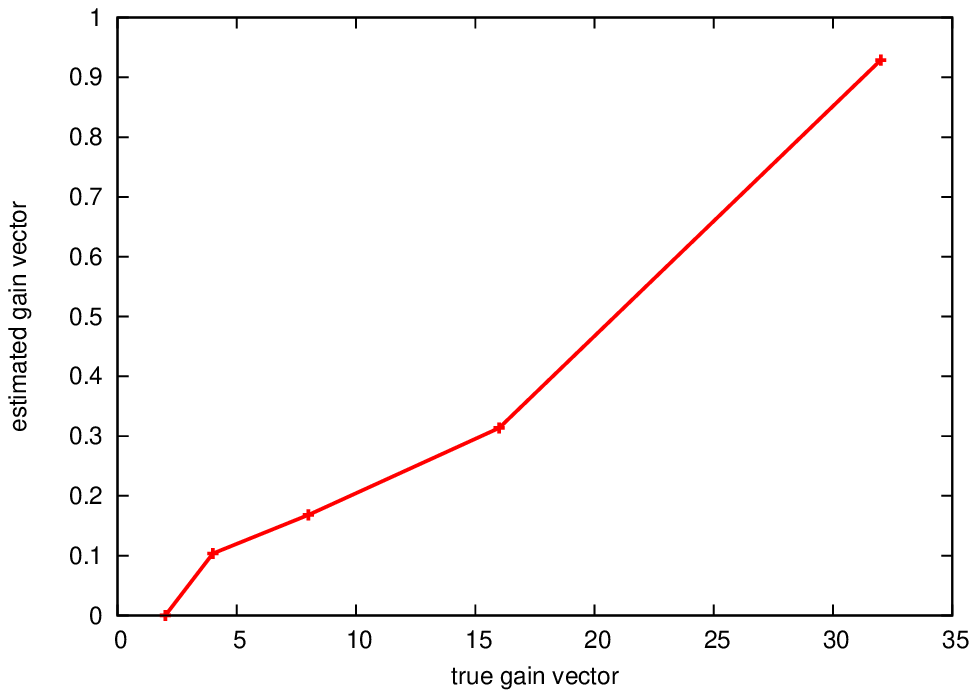}
\caption{The estimated gain vector with respect to the true gain vector} \label{fig:rankone_g}
\end{minipage}
\end{figure*}

After the utility function $w$ is obtained, we can reconstruct the gain vector and the discount factors from $w$. To this end, we rewrite $w$ as a matrix $W$ of size $L \times K$. Assume that singular value decomposition of the matrix $W$ can be expressed as $W = U{\rm diag}(\sigma_1, \dots, \sigma_n) V^T$ where $\sigma_1\geq\dots\geq\sigma_n$. Then, the rank-1 approximation of $W$ is $\hat W = \sigma_1u_1v_1^T$. In this case, the first left singular vector $u_1$ is the estimation of the gain vector and the first right singular vector $v_1$ is the estimation of the discount factors. We plot the estimated gain vector and discount factors with respect to their true values in Figure \ref{fig:rankone_g} and Figure \ref{fig:rankone_c}, respectively. Note the perfect estimations give straight lines in these two figures. We can see that the discount factors for the top-ranked positions are more close to a straight line and thus are estimated more accurately. This is because the discount factors of top-ranked positions have greater impact to the preference of the ranking lists. Therefore, these discount factors are captured more precisely by the constraints. The similar phenomenon can also be observed for the gain vector.

\subsection{Noisy Settings}
\begin{figure*}[t]
\centering
\begin{minipage}[t]{0.48\textwidth}
\includegraphics[width=1.0\textwidth]{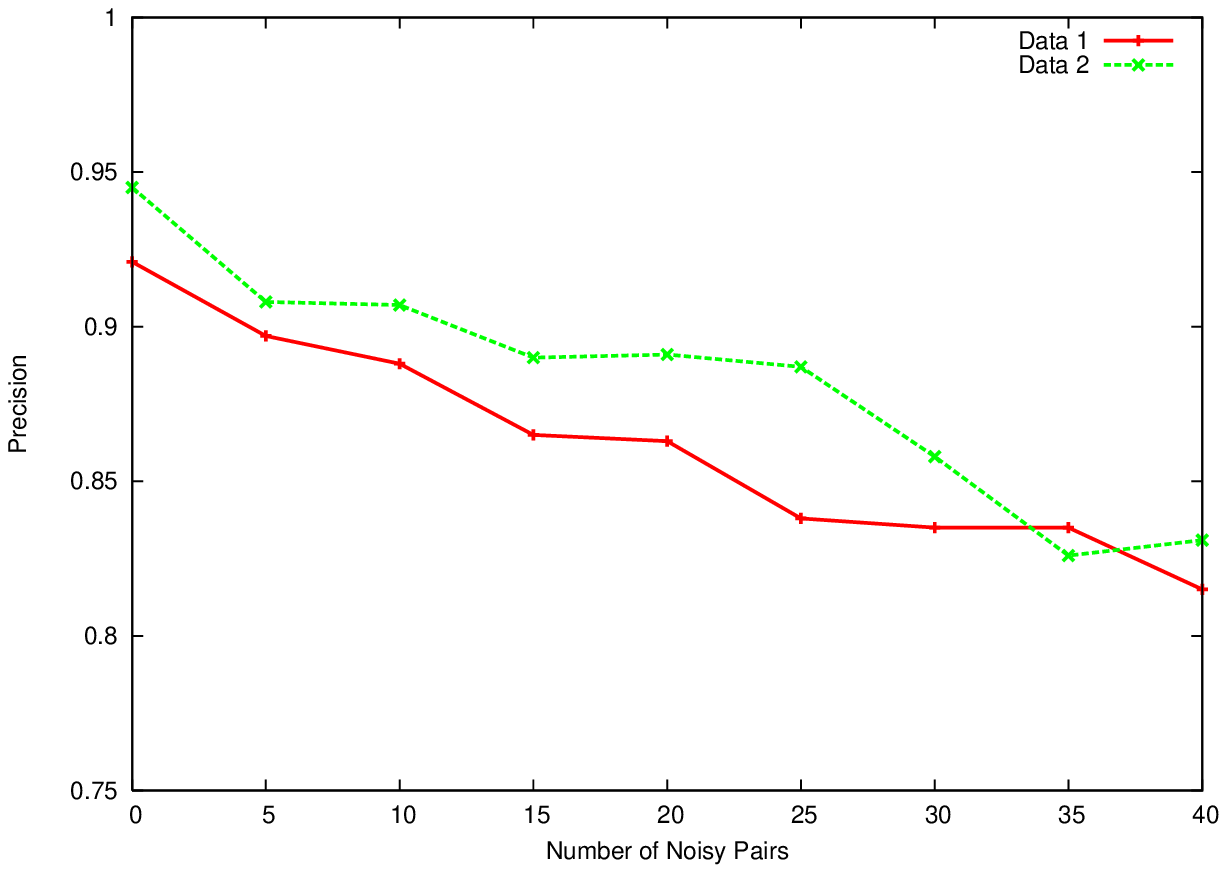}
\caption{Precision over the test set with respect to the number of noisy pairs in noisy settings} \label{fig:noise_p}
\end{minipage}
\quad
\begin{minipage}[t]{0.48\textwidth}
\includegraphics[width=1.0\textwidth]{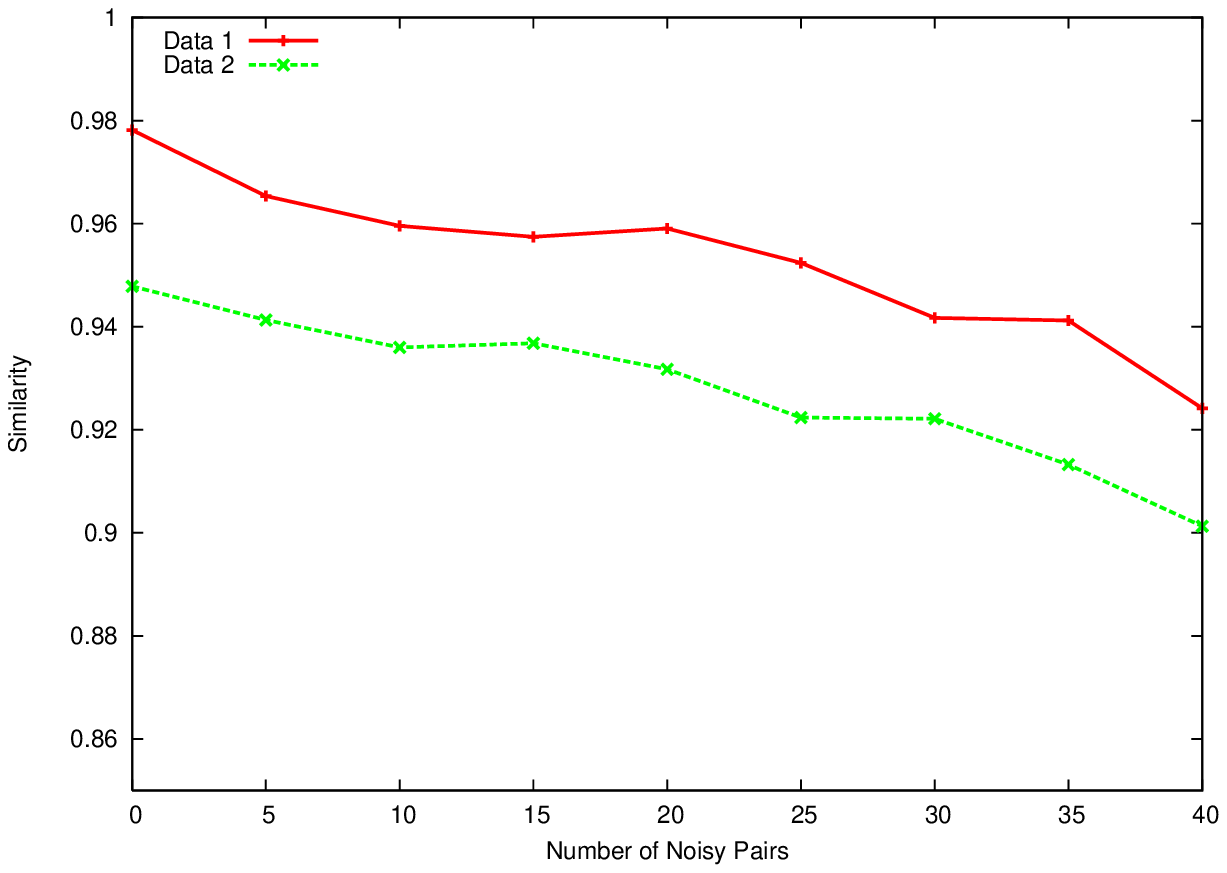}
\caption{Similarity between $w$ and $\hat w$ with respect to the number of noisy pairs in noisy settings} \label{fig:noise_sim}
\end{minipage}
\end{figure*}

In real world scenarios, the preference pairs of ranking lists can be noisy. Therefore, it is interesting to investigate the effect of the noisy pairs to the performance. To this end, we fix the number training pairs to be 200 and create the noisy pairs by randomly flip a number of pairs in the training set. In our experiments, the number of noisy pairs ranges from 5 to 40. Since the trade off value $C$ is important to the performance in the noisy setting, we select the value of $C$ that shows the best performance on an independent validation set. We  report the performance with respect to the number of noisy pairs in Figure \ref{fig:noise_p} and Figure \ref{fig:noise_sim}. We can observe that the performance decreases when the number of noisy pairs grows.

\begin{figure*}[t]
  \begin{minipage}[t]{0.48\textwidth}
  \includegraphics[width=1.0\textwidth]{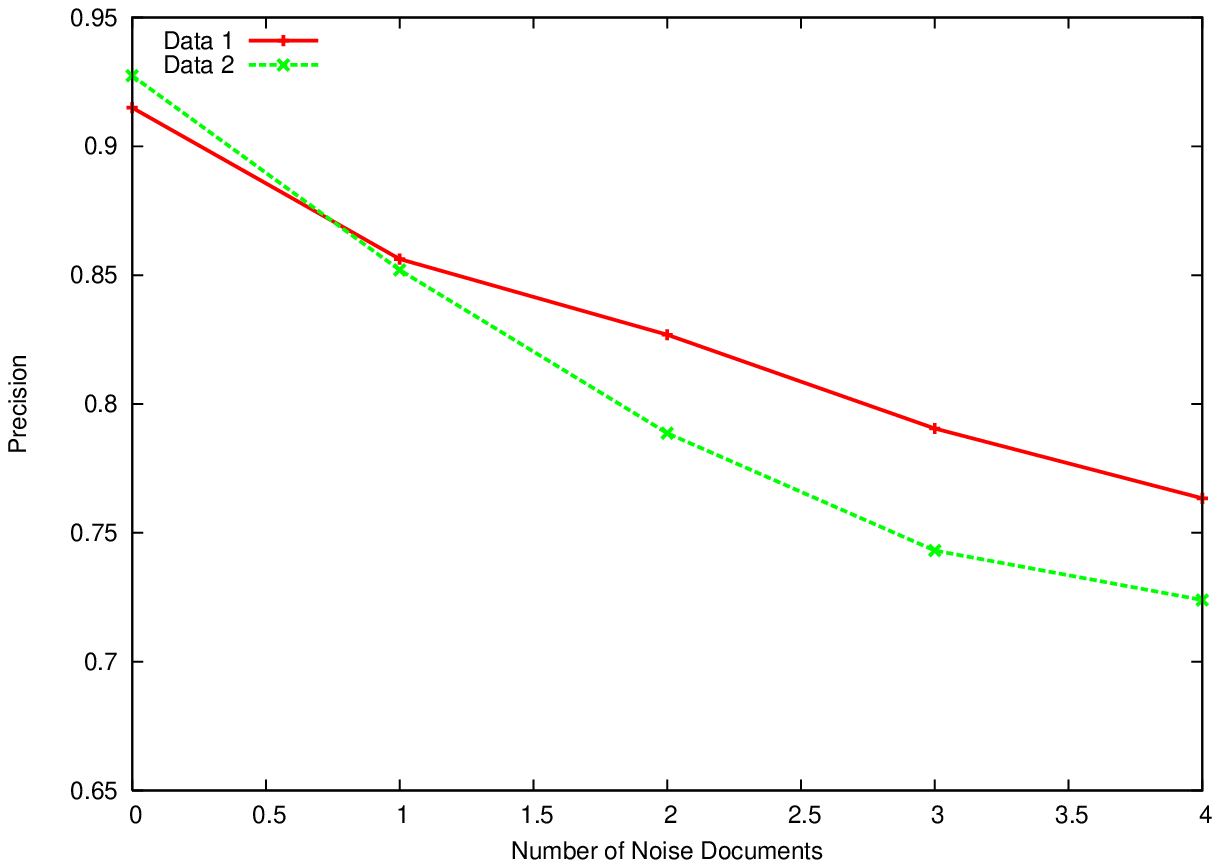}
  \caption{Precision with respect to the number of noisy grades in noisy settings} \label{fig:grade_noise_p}    
  \end{minipage}
  \quad
  \begin{minipage}[t]{0.48\textwidth}
  \includegraphics[width=1.0\textwidth]{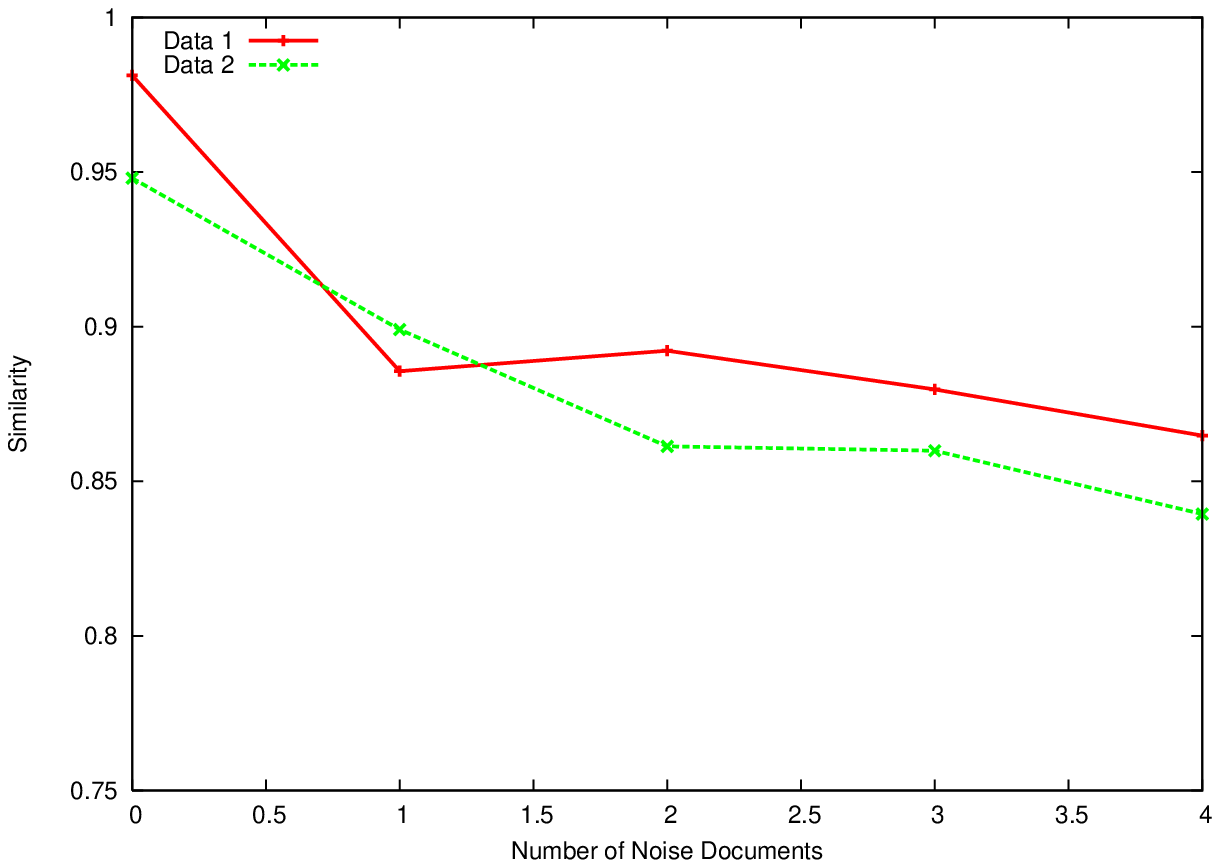}
  \caption{Similarity with respect to the number of noisy grades in noisy settings} \label{fig:grade_noise_sim}    
  \end{minipage}  
\end{figure*}

In addition to the noisy preference pairs, we also consider the noise in the grades of documents. In this case, we randomly modify the grades of a number of documents to form noise in training set. The estimated $\hat w$ is used to predict to preference on a test set. The precision with respect to the number of noisy documents is shown in Figure \ref{fig:grade_noise_p}. It can be observed that the performance decreases when the number of noisy documents grows.

\subsection{Optimal Rankings}
\begin{figure*}[htp]
\centering
\centering
\begin{minipage}[t]{0.48\textwidth}
\includegraphics[width=1.0\textwidth]{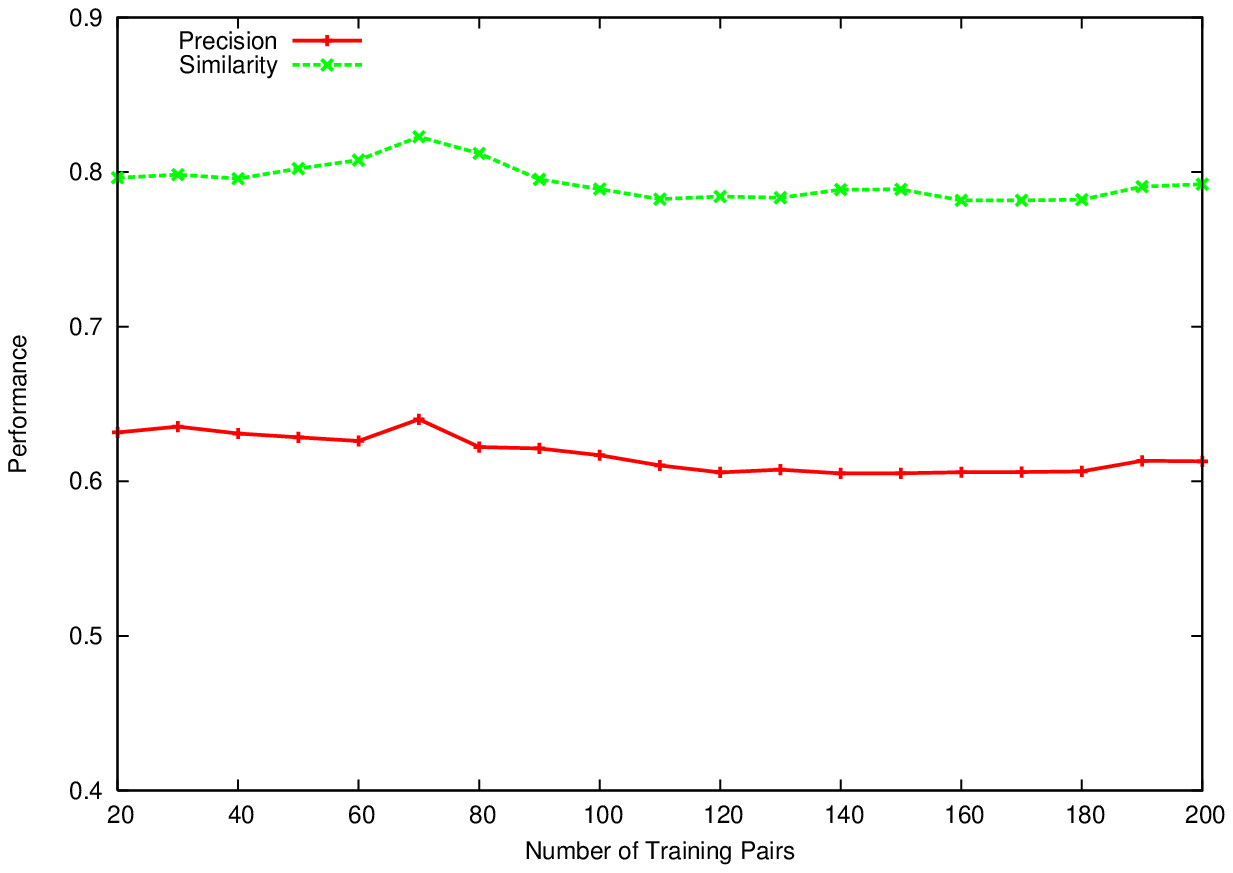}
\caption{Performance when training pairs are generated by permuting the same list} \label{fig:best1}
\end{minipage}
\quad
\begin{minipage}[t]{0.48\textwidth}
\centering
\includegraphics[width=1.0\textwidth]{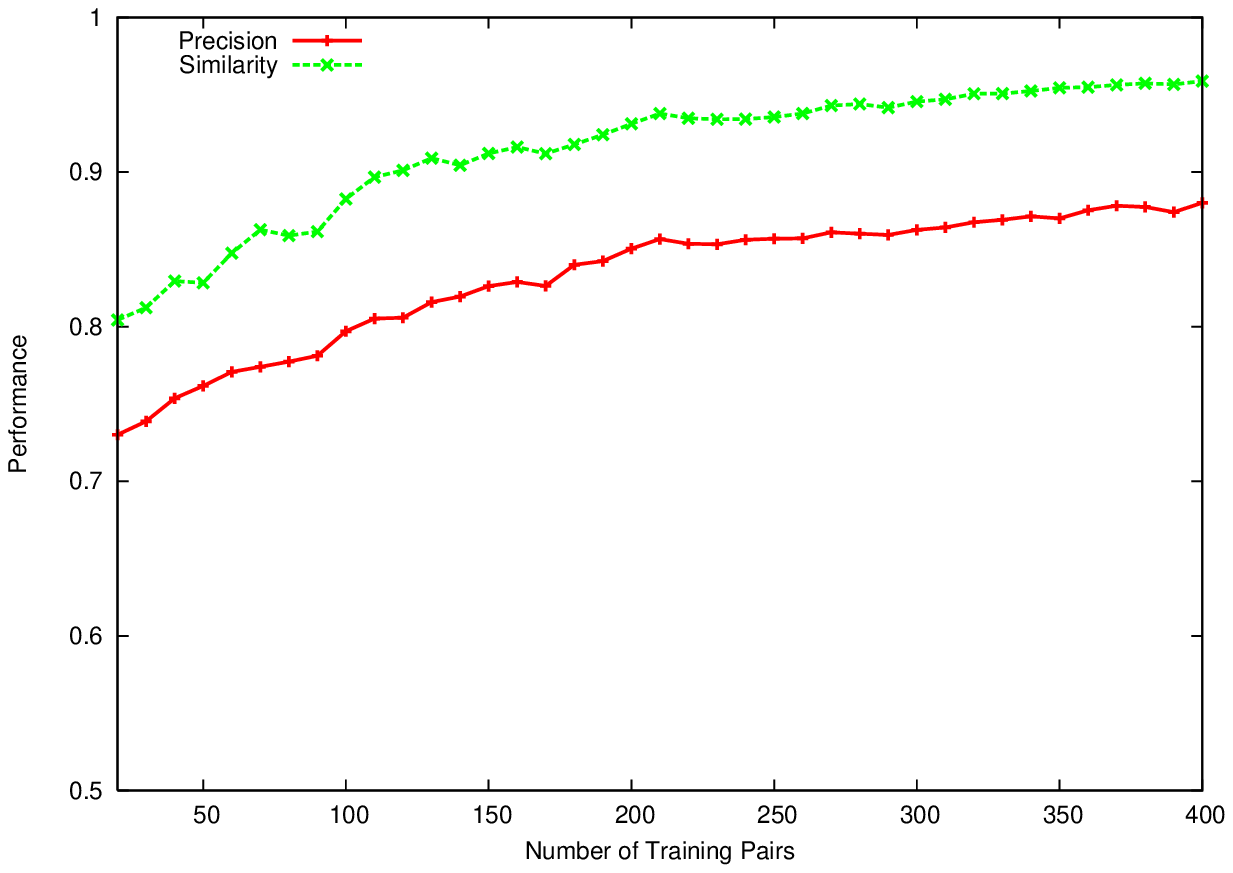}
\caption{Performance when training pairs are generated from different lists} \label{fig:best2}
\end{minipage}
\end{figure*}

We can further restrict the preference pairs by involving the optimal ranking in each pair of training data. For example, we set one ranking list of the preference pair to be the optimal ranking $[5,5,4,4,3,3,2,2,1,1]$.  In this case, if the other ranking list is generated by permuting the same list, it is implied by Proposition 1 that any compatible gain vectors will agree on the optimal ranking is preferred to other ranking lists. In other words, the preference pairs do not carry any constraints to the utility function $w$. Therefore, the constraints corresponding to these preference pairs are not effective in determining the utility function $w$. Consequently, the performance do not increase when the number of training pair grows as shown in Figure \ref{fig:best1}.

If the ranking lists contain different sets of grades, a fraction of constraints can be effective. The performance grows slowly with the number of training pairs increases as reported in Figure \ref{fig:best2}. By comparing Figure \ref{fig:simu_p},\ref{fig:simu_sim} and Figure \ref{fig:best2}, we can observe that when the type of preference is restricted, the learning algorithm requires more pairs to obtain a comparable performance. We conclude from this observation that some pairs are more effective than others to determine $w$. Thus, if we can design algorithm to select these pairs, the number of pairs required for training can be greatly reduced. How to design algorithms to select effect preference pairs for learning DCG will be addressed as a future research topic. 

\section{An enhanced model}
The objective function defined in Equation (\ref{eqn:obj1}) does not consider the degree of difference between ranking lists. For example, it deals with preference pairs $[5, 5, 4, 2, 1]\\ \succ [5, 4, 5, 2, 1]$ and $[5, 5, 4, 2, 1]\succ [2, 1, 5, 5, 4]$ in the same approach, although they have great differences in DCG. In order to overcome this problem, we propose an enhanced model that takes the degree of difference between ranking list into consideration.

\begin{eqnarray}
\min w^Tw + C\sum_{(i, j)\in S}\xi_{ij}^2
\end{eqnarray}
subject to:
\begin{eqnarray}
w^T(s(\pi_i)-s(\pi_j))\geq \textrm{Dist}(\pi_i, \pi_j)-\xi_{ij} \quad (i, j)\in S\nonumber\\
\xi_{ij}\geq 0 \quad (i, j)\in S\nonumber\\
w_{kl}\geq w_{k,l+1} \quad  k=1, \dots, K \textrm{ and } l = 1, \dots, L-1\nonumber
\end{eqnarray}
where $\textrm{Dist}()$ is a distance measure for a pair of permutations $\pi_1$ and $\pi_2$. In principle, we would prefer $\textrm{Dist}(\pi_i, \pi_j)$ to be a good approximation of $DCG(\pi_i)-DCG(\pi_j)$. However, since we do not actually know the ground-truth $w$ in practice, it is generally difficult to obtain a precise approximation. In our simulation, we apply the Hamming distance as the distance measure:
\begin{eqnarray}
\textrm{Ham}(\pi_1, \pi_2) = \sum_k {1}[\pi_1(k)\neq\pi_2(k)]
\end{eqnarray}

\begin{figure*}[ht]
\centering
\begin{minipage}[t]{0.48\textwidth}
\includegraphics[width=1.0\textwidth]{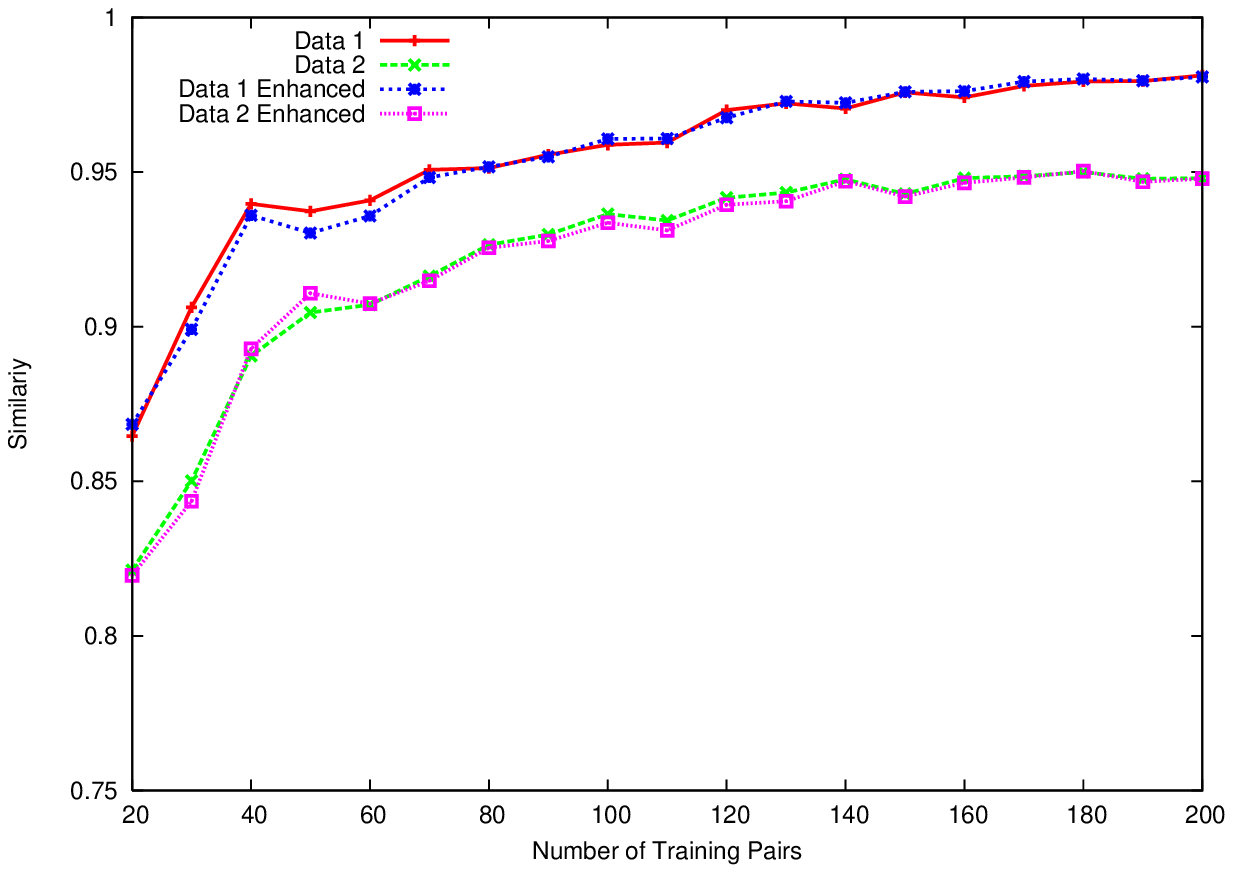}
\caption{Similarity of the enhanced model with respect to the number of training pairs} \label{fig:hamming_sim}
\end{minipage}
\quad
\begin{minipage}[t]{0.48\textwidth}
\center
\includegraphics[width=1.0\textwidth]{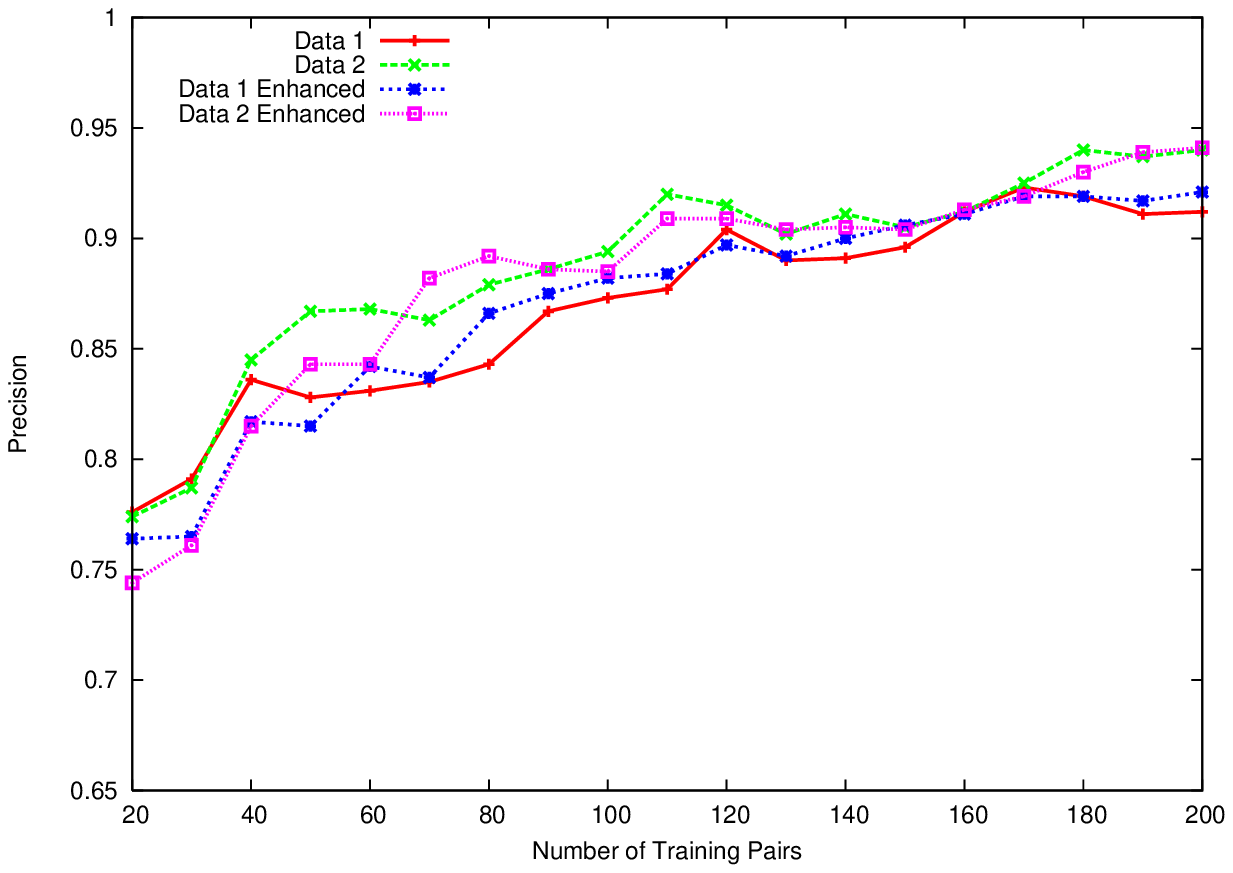}
\caption{Precision of the enhanced model over the test set with respect to the number of training pairs} \label{fig:hamming_p}
\end{minipage}
\end{figure*}

We perform simulation to evaluate the enhanced model. Form Figure \ref{fig:hamming_p} and Figure \ref{fig:hamming_sim}, we can observe that the performance improvement of the enhanced model is not very significant.  We doubt that this is because Hamming distance is not a precise approximation of the DCG difference. We plan to investigate this problem in our future study.

\section{Learning without document grades}
When the grades of the documents are not available, we can also fit a model to predict the preference of two ranking list. To this end, we use a $K\times K$-dimensional binary vector $s(\pi)$ to represents the ranking $\pi$. The first $K$ components of $s(\pi)$ correspond to the first position of the K-position ranking, and the second $K$-components the second the position, and so on.

\begin{figure}[t!h]
\centering
\includegraphics[width=0.5\textwidth]{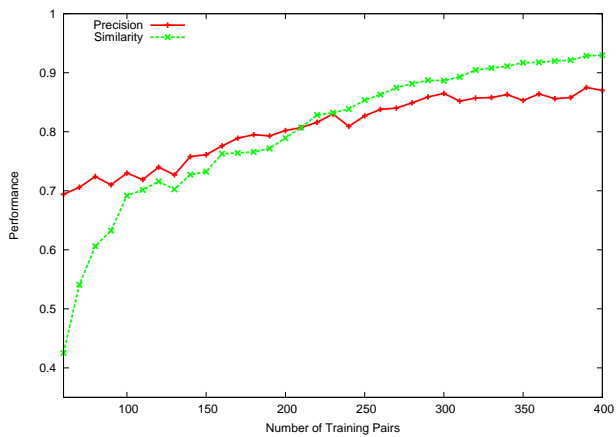}
\caption{Performance over the test set with respect to the number of training pairs when the grades of documents are not known} \label{fig:docrank}
\end{figure}

For example, for a ranking list $\pi$
$$
d_3, d_1, d_4, d_2, d_5
$$
the corresponding binary vector is
$$
[0, 0, 1, 0, 0,\;\;1, 0, 0, 0, 0,\;\;0, 0, 0, 1, 0,\;\;0, 1, 0, 0, 0,\;\;0, 0, 0, 0, 1]
$$
Given a set of preference over ranking lists, we can obtain $w$ by solving the following optimization problem:
\begin{eqnarray}
\min w^Tw + C\sum_{(i, j)\in S}\xi_{ij}^2
\end{eqnarray}
subject to:
\begin{eqnarray}
w^T(s(\pi_i)-s(\pi_j))\geq 1-\xi_{ij} & (i, j)\in S\\
\xi_{ij}\geq 0 & (i, j)\in S
\end{eqnarray}
In this case, the constraints $w_{kl}\geq w_{k,l+1}$ are not included in the optimization problem, since we do not have any prior knowledge about the grades of the documents. The precision on test set with respect to the number of training pairs is reported in Figure \ref{fig:docrank}. We can observe that the $w$ can be precisely learned even without the grades of documents. In this case, the learned utility function $w$ can be interpreted as the relevant judgements for the documents.

\section{Conclusions and Future Work}
In this paper, we investigate the coherence of  DCG, which is an important performance measure in information retrieval. Our analysis show that the DCG is incoherency in general, i.e., different gain vectors can lead to different judgements about the performance of a ranking function. Therefore, it is a vital problem to select reasonable parameters for DCG in order to obtain meaningful comparisons of ranking functions. We propose to learn the DCG gain values and discount factors from preference judgements of ranking lists. In particular, we develop a model to learn DCG as a linear utility function and formulate the method as a quadratic programming problem. Preliminary results of simulation suggest the effectiveness of the proposed method.

We plan to further investigate the problem of learning DCG and apply the proposed method in real world data sets. Furthermore, we plan to generalize DCG to nonlinear utility functions to model more sophisticated requirements of ranking lists, such as diversity and personalization.

\bibliographystyle{abbrv}
\bibliography{dcg}

\begin{thebibliography}{1}

\bibitem{jmr1997:carrol}
J.~D. Carroll and P.~E. Green.
\newblock {P}sychometric methods in marketing research: {P}art {I}{I},
  {M}ultidimensional scaling.
\newblock {\em Journal of Marketing Research}, 34:193--204, 1997.

\bibitem{nips2004:chapelle}
O.~Chapelle and Z.~Harchaoui.
\newblock A machine learning approach to conjoint analysis.
\newblock volume~17, pages 257--264, Cambridge, MA, USA, 2005. MIT Press.

\bibitem{ms2005:evgeniou}
T.~Evgeniou, C.~Boussios, and G.~Zacharia.
\newblock Generalized robust conjoint estimation.
\newblock {\em Marketing Science}, 24(3):415--429, 2005.

\bibitem{sigir2000:jarvelin}
K.~J\"{a}rvelin and J.~Kek\"{a}l\"{a}inen.
\newblock Ir evaluation methods for retrieving highly relevant documents.
\newblock In {\em SIGIR '00: Proceedings of the 23rd annual international ACM
  SIGIR conference on Research and development in information retrieval}, pages
  41--48, New York, NY, USA, 2000. ACM.

\bibitem{icml2005:joachims}
T.~Joachims.
\newblock A support vector method for multivariate performance measures.
\newblock In {\em Proceedings of the 22nd international conference on Machine
  learning}, 2005.

\bibitem{book:louviere}
J.~J. Louviere, D.~A. Hensher, and J.~D. Swait.
\newblock {\em Stated choice methods: analysis and application}.
\newblock Cambridge University Press, New York, NY, USA, 2000.

\bibitem{sigir01:voorhees}
E.~M. Voorhees.
\newblock Evaluation by highly relevant documents.
\newblock In {\em SIGIR '01: Proceedings of the 24th annual international ACM
  SIGIR conference on Research and development in information retrieval}, pages
  74--82, New York, NY, USA, 2001. ACM.

\bibitem{sigir2007:xu}
J.~Xu and H.~Li.
\newblock Adarank: a boosting algorithm for information retrieval.
\newblock In {\em Proceedings of the 30th ACM SIGIR}, pages 391--398, New York,
  NY, USA, 2007.

\bibitem{sigir07:yue}
Y.~Yue, T.~Finley, F.~Radlinski, and T.~Joachims.
\newblock A support vector method for optimizing average precision.
\newblock In {\em Proceedings of ACM SIGIR}, New York, NY, USA, 2007.

\end{thebibliography}
\end{document}